# Electrokinetic Flow Modeling Through Bone Scaffold

Mohammad Osaid[1], Debabrata Dasgupta[2]

[1] KTH Royal Institute of Technology, Stockholm, Sweden
[2] Indian Institute of Technology Delhi, India

## Abstract

Fluid flow and mass transfer inside a bioreactor play a pivotal role in growing bone grafts, as cell proliferation is limited by the transport of nutrients and oxygen, as well as the removal of by-products from cells within the scaffold. Traditionally, perfusion bioreactors are used for tissue-engineered bone grafts. In this study, we modeled electrokinetic flow through the graft for tissue-engineered bone grafts and compared it to pressure-driven flows. The study highlights that electrokinetic transport offers several advantages over pressure-driven flow, such as improved oxygen transport and higher shear stress. Additionally, it provides generic benefits, including greater flexibility in control mechanisms, such as easier actuation and manipulation of flow.

## Introduction

Fluid flow and mass transport through bone scaffolds play a critical role in controlling bone graft growth for clinical applications. Effective delivery of oxygen and glucose is essential for cellular metabolism and proliferation, as well as the efficient removal of waste products generated by cellular activities. Accumulation of these metabolic by-products inhibits cell growth, highlighting the necessity for an optimized transport mechanism within the scaffold. The transport of metabolites represents a fundamental limiting factor in determining the size and viability of tissue-engineered bone grafts.

To facilitate metabolite transport and support cellular activities, scaffolds are typically cultured in dynamic environments such as bioreactors. Traditional static culture systems, such as Petri dishes, rely solely on diffusive transport, which restricts nutrient penetration to the scaffold's central regions, thereby impeding tissue growth. Dynamic bioreactors, on the other hand, leverage fluid dynamics to enhance metabolite transport and support uniform cell proliferation throughout the scaffold. Numerous studies have demonstrated the superiority of dynamic systems over static ones [1, 2]. For instance, spinner flasks—a type of non-static bioreactor—employ stirring mechanisms to enhance metabolite transport and induce shear stress on scaffold surfaces, which significantly influences tissue growth[3, 4] . Vunjak et al.[5] conducted comparative analyses of cartilage regeneration in spinner flasks and static systems, revealing superior cell proliferation in the former.

Perfusion bioreactors offer additional advantages by combining convection and diffusion for enhanced metabolite transport and efficient waste removal. Studies by Goldstein et al. and Grayson et al. demonstrated improved bone formation by mesenchymal stem cells (MSCs) cultured in perfusion bioreactors compared to static systems[6-8] . While shear stress in these systems enhances cellular proliferation and differentiation, excessive magnitudes can also adversely affect cell growth[9]. Whittaker et al. [10] observed reduced osteoblastic cell growth under high shear stress, underscoring the importance of optimizing stress levels in bioreactor designs.

This study models electro-kinetic transport in bone scaffolds placed in a perfusion bioreactor, diverging from conventional pressure-gradient-induced flows. Instead, an electric field is employed to drive fluid flow, fundamentally altering the transport physics and bioreactor design. The scaffold, derived from the decellularized trabecular bone of bovine femurs, was analyzed for the transport of oxygen and glucose under applied electric fields across its cubic geometry (2 mm per side).

# Theoretical Framework and Setup

### Background

This study focuses on the electro-kinetic transport of electrolytes, specifically oxygen, through bone scaffolds with length scales of a few centimeters or more. It examines key parameters such as wall shear stress, oxygen concentration, and velocity. Electrokinetic transport is caused by the polar interactions between solid and liquid phases. These interactions create an electrical double layer on surfaces, accelerating fluid flow under an external electric field.

Trabecular bone scaffolds consist of two regions: the solid domain and the liquid domain. The liquid domain is a void volume where Newtonian fluid, carrying dissolved oxygen, flows and transports nutrients. The solid domain blocks fluid convection and allows only diffusive transport of electrolytes, which are consumed by cells in the bone following Michaelis–Menten kinetics. Thus, Navier-Stokes equations are solved for the liquid domain, while diffusion equations are solved for the solid domain. However, solving these equations for large, complex structures like bones is computationally expensive and prone to inaccuracies due to the intricate geometry.

To address these challenges, a smaller portion of the bone, known as the Reference Volume Element (RVE), is used for simulations, similar to the previous study [11]. The RVE captures the detailed geometry of the bone, reducing computational costs and improving solution accuracy. Once the governing equations are solved for the RVE, the results are upscaled using homogenization techniques to represent the larger bone sample. This method ensures efficient and accurate modeling across different length scales.

### Governing Equations

This section outlines the governing equations for electro-kinetic transport through the RVE and their upscaling using the homogenization technique. The RVE length scale (l) is much smaller than the macro sample length L, ensuring $\varepsilon = \frac{l}{L} \ll 1$, which allows the separation of scales required for homogenization. The RVE, as shown in Fig. 1-b, consists of solid and liquid domains, where polar interactions induce surface charges, forming an Electric Double Layer (EDL) with a constant zeta potential ($\zeta$) on the surfaces.

To compute the induced electric potential ($\psi$), we use the Poisson equation:

$$\nabla^2 \psi = -\rho_e/\varepsilon_r \qquad (1)$$

where $\rho_e$ is the induced charge, and $\varepsilon_r$ is the permittivity of the media. Assuming $\varepsilon_r$ is constant, $\rho_e$ is related to ion concentrations via the Nernst equation:

$$\rho_e = e\,(z^+ n^+ + z^- n^-) \qquad (2)$$

where e is the electronic charge, $z+$ and $z-$ are ion valencies, and $n^+$, $n^-$ are ion concentrations. The ion concentrations follow the Boltzmann distribution:

$$n_\pm = n_0 \exp(-z^\pm e\psi/k_B T) \qquad (3)$$

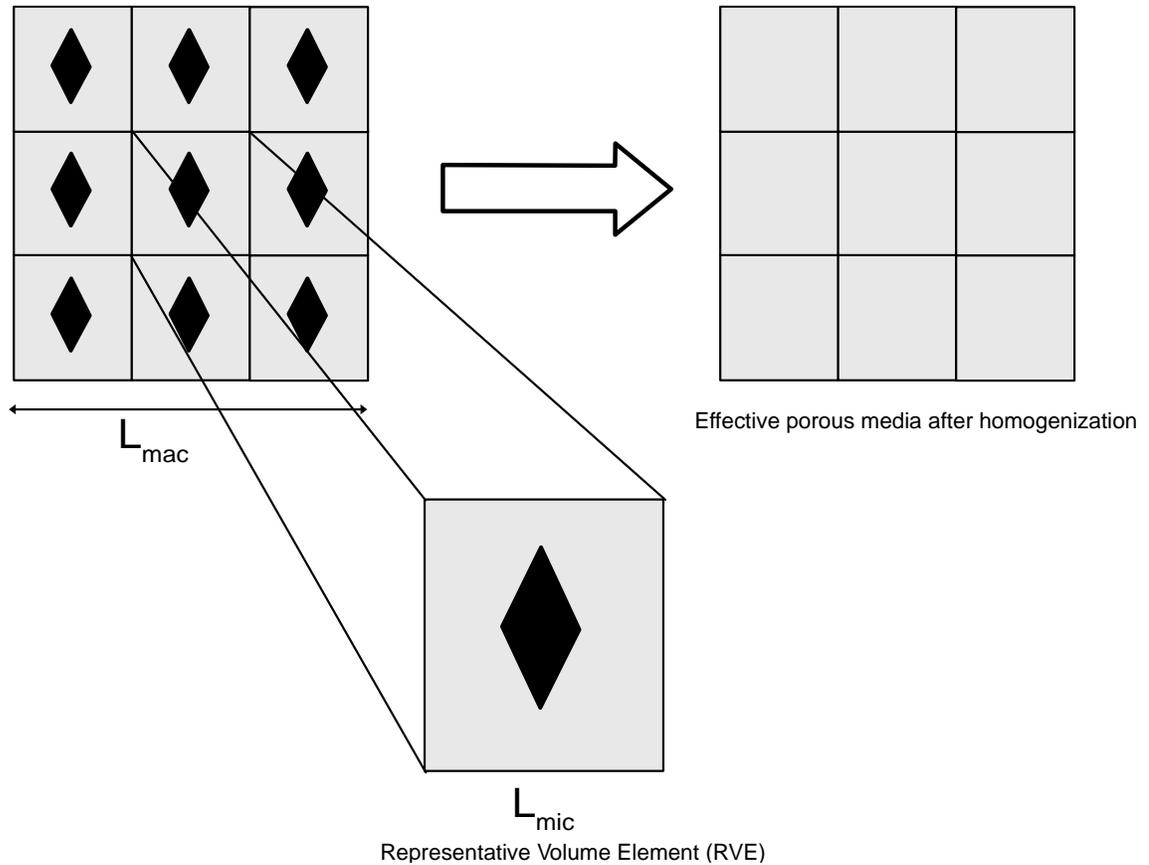

**Fig 1**: Schematic of the spatially periodic cells. Representative volume elements (RVEs or REVs) are periodic and serve as repeating units to form the entire structure. The black in the center is solid domain whereas rest is liquid domain.

Combining the Poisson and Boltzmann equations gives:

$$\nabla^2 \psi = (2zen_0/\varepsilon_r)\,\sinh(ze\psi/k_B T) \qquad (4)$$

This equation is solved under periodic boundary conditions with a zeta potential ($\zeta$) at the walls.

An external electric field ($\phi$) is applied to drive fluid flow, satisfying:

$$\nabla^2 \varphi = 0 \qquad (5)$$

The governing equations for convection and diffusion of oxygen are:

Continuity Equation: $\quad \nabla \cdot v = 0 \qquad (6)$

Naviers Stoke: $\quad \rho \vec{v} \cdot \vec{\nabla} \vec{v} = -\vec{\nabla} p + \mu \nabla^2 \vec{v} + \rho_e \vec{E} \qquad (7)$

Diffusion Equation: $\quad c \dfrac{\partial u}{\partial x} + D \dfrac{\partial^2 u}{\partial x^2} = 0 \qquad (8)$

Where v is velocity, ρ is density, p is pressure, µ is viscosity, E is applied electric field

The additional term in the right hand side of the Naviers Stoke equation represents the forces due to the electrical interaction. Likewise the above cases the walls are assumed to be periodic for solving these equation along with no slip and no flux boundary condition at the walls.

Here, v is velocity, ρ is density, p is pressure, µ is viscosity, E is the applied electric field, c is concentration, and D is the diffusion coefficient. The electrical force term in the Navier-Stokes equation accounts for interactions due to the EDL.

All equations are solved with periodic boundary conditions on the walls, except for the inlet and outlet, where constant values are applied. No-slip and no-flux conditions are also imposed at the walls.

The governing equations are first non-dimensionalized to simplify the analysis and ensure the equations are scale-independent. The parameters are then expanded asymptotically to enable the homogenization process, as described in this study [11]. This leads to the following dimensionless equation:

$$\nabla \bar{\psi_0} = (A^2/\zeta) \sinh(\zeta \bar{\psi_0}) \qquad (8)$$

$$A = L_{mic}/\lambda \qquad (9)$$

$$\lambda = \mathrm{sqrt}(\,(\varepsilon_{r,0}\, k_B\, T) / (2\, z^2\, e^2\, n_0)\,) \qquad (10)$$

**Setup**

To prepare the geometry of the trabecular bone for the numerical study, a small cancellous sample was taken from the tibia of a 9-12-month-old bovine. The sample was first cleaned using a water jet to remove cellular material and then cut into a smaller cube, a few millimeters in size, using a vibration cutter to avoid distortion of the bone's microstructure. A micro-CT scanner was used to generate 2D slices of the bone with dimensions 2 × 2 mm (Figure 2a). These images were processed using Scan IP software to create a 3D cubical bone mesh of size 2 × 2 × 2 mm³. This mesh, shown in Figure 2 b, represents the Reference Volume Element (RVE) of a larger bone sample.

All governing equations were solved for the RVE, and the results were upscaled using the homogenization technique. This approach ensures an accurate representation of the larger bone sample, provided the RVE is significantly smaller than the full sample, typically in the centimeter scale or larger. Using a smaller RVE is avoided, as it may fail to capture the bone's topographical details accurately.

# Results

## Pressure-based flow on simplified geometry

The simplified geometry was first used to model the scenario where the geometry consists of a periodic cubic structure with spherical holes designed to allow fluid flow. A porous media approach

a)

b)

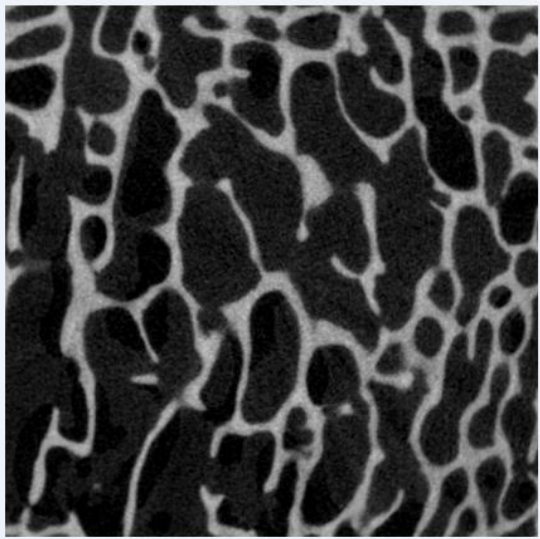 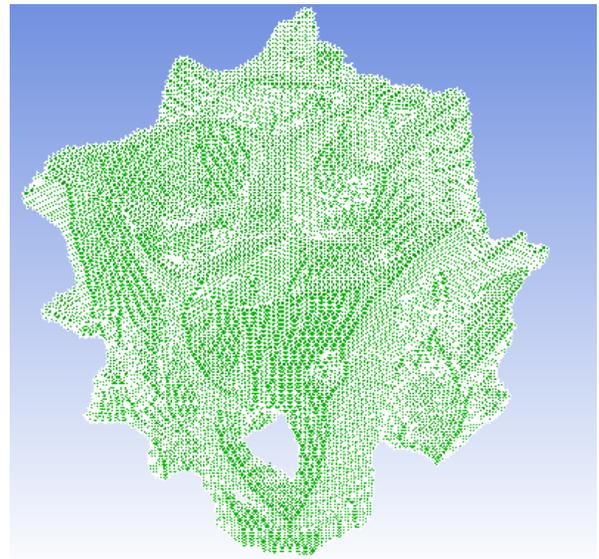

**Fig 2**: a) Image of a layer obtained using μ-CT Scan. b) Solid domain of the trabecular bone.

applied in the simulation, where the fluid experiences resistive forces, and the momentum conservation equation is used to model the flow. The resistive force is calculated based on the permeability of the material, which can be determined experimentally using Darcy's Law. For this study, the permeability is assumed to be $4.5 \times 10^{-10}$. Since the Reynolds number is low, inertial forces are neglected. Water is used as the working fluid due to its similarity in density and viscosity to DMEM, which is commonly used in biological applications. The inlet velocity is set to 0.05 m/s, while the outlet is modeled as a constant pressure boundary. No-slip boundary conditions are applied to the walls, ensuring zero fluid velocity at the surface, while symmetric boundary conditions are used for other faces (Figure 3).

Upon convergence of the simulation, pressure, velocity, and wall shear stress contours are analyzed. As the fluid flows through the porous structure, resistive forces cause a pressure drop. However, excessive velocity, while increasing pressure and aiding diffusion, can adversely affect cells. Wall shear stress is also critical, as it directly impacts cell growth. Therefore, optimizing flow conditions to balance diffusion and mechanical stress is essential for applications involving fluid-cell interactions.

The variation of stress versus velocity is studied; as the velocity of the nutrient media increases, the shear stress on the wall of bone also increases. Using this simulation, the velocity of fluid could be optimized to grow a bone or keep the bone cells alive. The variation in shear stress with velocity is discussed in the next section (Figure 4a). Oxygen concentrations inside the scaffold were simulated, considering two regions: the bulk medium phase and the cell-layer region. Transport in the bulk medium phase was modeled using the Navier–Stokes and convection-diffusion equations, while the cell-layer region was modeled using only the convection-diffusion equation.

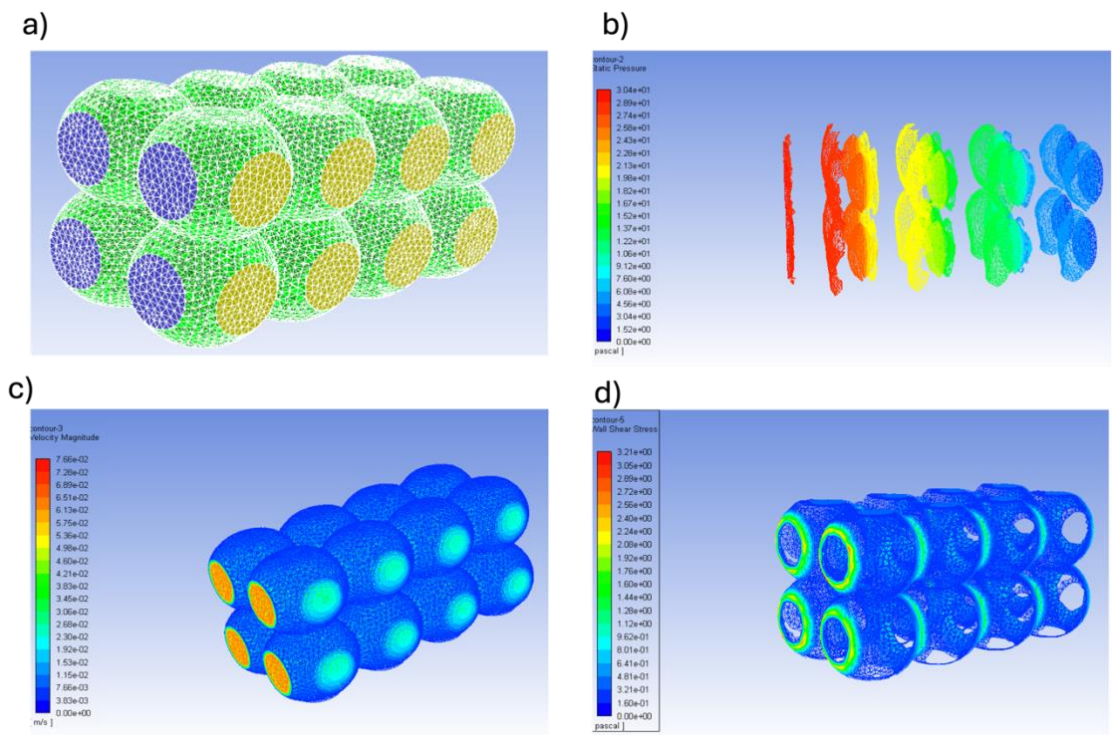

**Fig 3**: Result of pressure-based flow on a simplified geometry. a) Flow domain with suitable boundary conditions. b) Pressure distribution inside the sample. c) Velocity Contour inside sample. d) Shear stress on the wall

The governing convection-diffusion equation is:

$$\partial c/\partial t = D\nabla^2 c - u \cdot \nabla c + R$$

Where:
- c is the oxygen concentration (mol/m³),

- D is the diffusion coefficient (3.29 × 10⁻⁹ m²/s in the medium region and 2.0 × 10⁻⁹ m²/s in the cell layer),
- R is the oxygen consumption rate.
- The oxygen consumption rate R follows Michaelis–Menten kinetics:

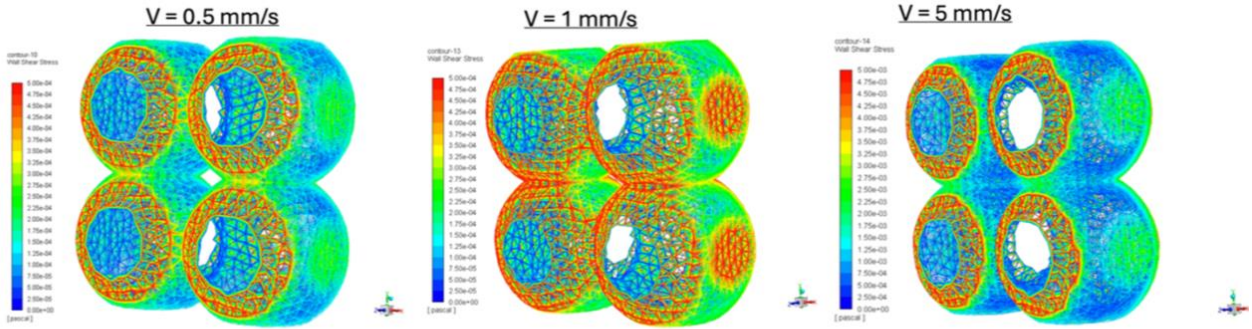

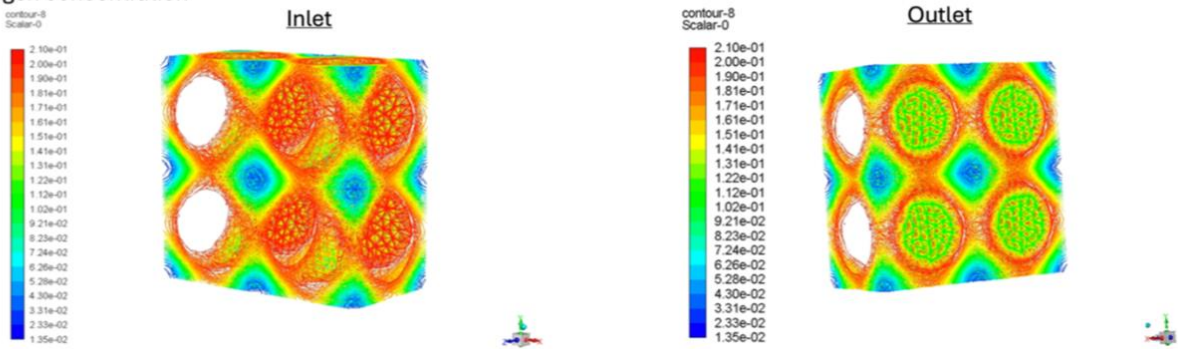

**Fig 4:** Results of pressure-based flow on a simplified geometry. a) Variation in shear stress on the wall. b) Oxygen concentration in a different parts of the scaffold.

- $R = -(1/v_{aell}) \times (Q_m c / (C_m + c))$

Where:

- $v_{aell}$ is the cell volume (1.44 × 10⁻¹⁵ m³/cell),
- $Q_m$ is the maximal oxygen consumption rate (1.86 × 10⁻¹⁸ mol/cell/s),
- $C_m$ is the oxygen concentration at half-maximal consumption (6.0 × 10⁻³ mol/m³).

The dissolved oxygen concentration at the inlet was assumed to be in equilibrium with air ($C_o$ = 2.1 × 10⁻¹ mol/m³)

The concentration of oxygen inside the bone scaffold was studied with variations in inlet and outlet oxygen concentrations. The results show that the middle part of the scaffold has the minimum oxygen concentration compared to the inlet and outlet regions (Figure 4b).

**Electrokinetic Transport in a 2D Domain**

For simplicity, electrokinetic transport is first modeled in a two-dimensional flow domain consisting of a circular solid embedded in a rectangular fluid region. The nutrient media flow is confined to the fluid domain, while oxygen diffuses through both the fluid and solid domains due to its diffusive transport properties. To drive the flow, a unit external electric field is applied across the domain, and its distribution is computed, as shown in Figure 5. The polar interaction between the nutrient media and the solid (representing bone) leads to the formation of an electrical double layer (EDL). The potential within this EDL is calculated by assigning a unit potential to the central solid and the walls of the flow domain. The resulting potential field shows a value of one within the circular region and on the side walls, gradually varying between zero and one in the rest of the domain.

The applied electric field induces ion motion, which in turn drags the nutrient media, creating flow. The figure 5 illustrates that the velocity field gradient in this case is higher compared to pressure-driven flow. Oxygen concentration is modeled by assigning an inlet concentration of 0.21 and an outlet concentration of 0.1. Oxygen consumption by bone cells for metabolic activities is incorporated into the model using a negative source term in the solid domain. This approach effectively captures the interplay between oxygen transport and cellular consumption within the system.

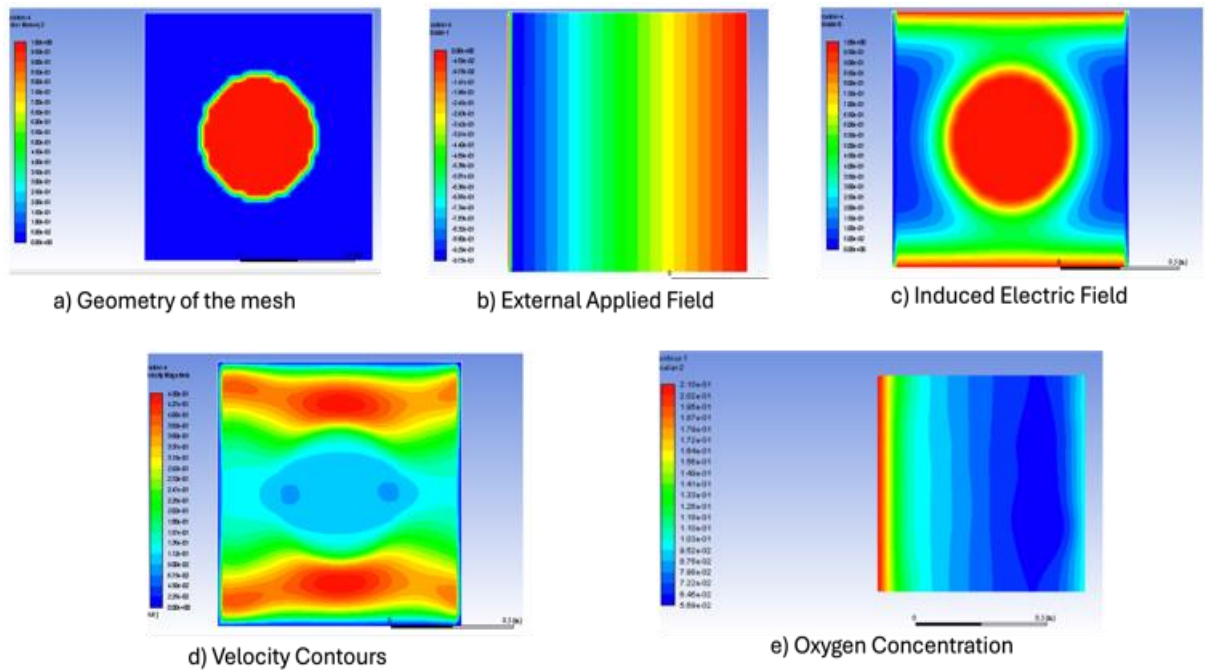

**Fig 5**: Results of electrokinetic flow in simplified 2D geometry.

**Electrokinetic Transport in bone mesh**

After validating the model on a simplified two-dimensional flow domain, it was extended to an actual bone mesh obtained from microCT scans of trabecular bone. The simulation results for the applied electric field, electrical double layer (EDL) field, velocity distribution, and oxygen concentration on the bone structure are shown in the Figure 6.

# Discussion

Two transport processes having completely different physics are chosen and compared; one is the pressure-driven flow, mainly used in commercial bioreactors, and the other is electrokinetic flow, where the electric field induces the flow. Both of these mechanisms are used to model the transport of a medium of nutrients, which has dissolved oxygen, through a trabecular bone mesh. In the former mechanism, a unit pressure gradient is applied in one direction after making the inlet and outlet periodic so as to generalize the results on the larger sample of bone, and all the other faces of the mesh are marked as a wall having no-slip boundary conditions. In electrokinetic transport, the polar interaction of ions and the surrounding media is first calculated in terms of electric double-layer potential. Further, a unit external electric field is applied across the bone, which accelerates the ion inside the nutrients media, which instigates the motion of the fluid through the bone and delivers the nutrients and oxygen to the cell. There are a couple of hydrodynamic and concentration factors that affect the growth and life of the bone cells. All these variables are compared in both transport processes to determine the better one.

In the given setup, the oxygen gets dissolved in the nutrient medium according to Henry's Law and convected through the bone. Apart from convective transport, it also diffuses through the scaffold. The concentration of oxygen is almost the same in both the transport processes in spite of having completely different physics behind the transport. The reason for the similar concentration is the low velocity, which substantially decreases the convective part of the transport and also affects other parameters. Apart from the concentration of oxygen in the scaffold, the velocity of the fluid also plays a crucial role. With the same pressure and voltage gradient applied across the bone, i.e., unit pressure gradient and unit electric field, the velocity in the case of electrokinetic transport is considerably higher than the other. Higher velocity in the case of electrokinetic transport increases the flux of nutrient medium through bone, which ensures better transport of glucose to the cells and hence increases cellular activity.

Apart from the above-mentioned parameters, cellular activity is also affected by the mechanical stimulation of the cells; the higher the mechanical stimulation, the more cell growth. So, like other variables, it is important to consider mechanical stimulation or shear stress in our discussion. Shear stress on the wall is proportional to the viscosity of the fluid and the velocity gradient. Hence, the velocity gradient is higher in electrokinetic transport because of the higher velocity of the fluid at the unit-applied field. The higher velocity causes large shear stress on the wall as compared to pressure-based transport, which has lower fluid velocity. An interesting thing in this discussion is the velocity profile; the velocity profile in the case of pressure-based flow is parabolic, whereas it is almost constant in the case of electrokinetic flow. The gradient of velocity is also dependent on the velocity profile; it is smaller in the case of a parabolic profile, whereas it is higher in a uniform flow profile.

From this, we can infer that the bone senses higher stress in electrically driven flow rather than pressure-driven.

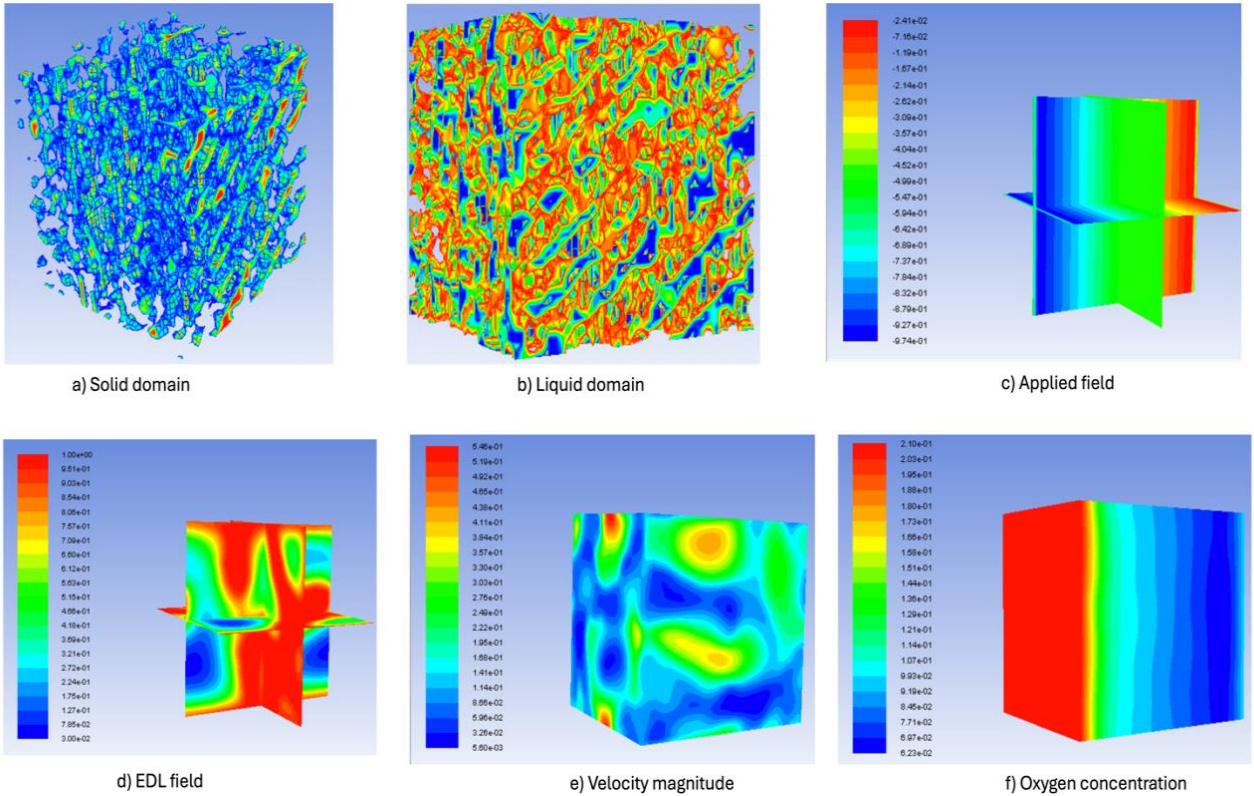

**Fig 6**: Results of electrokinetic flow in 3D bone mesh.

## Conclusion

This study indicates that the electrokinetics-based bioreactor for bone growth has several advantages over the pressure-based. Identical reactors having a unit potential and pressure gradient applied across it, the reactor having unit potential gradient have better transport of nutrients medium through the bone. Moreover, the pressure based reactor having unit pressure gradient have low shear on the wall which reduces the cellular activity. Apart from these edges, electrokinetic transport system have some generic advantages over its counterpart i.e it has higher flexibility in control mechanism which includes easy actuation and manipulation of flow.

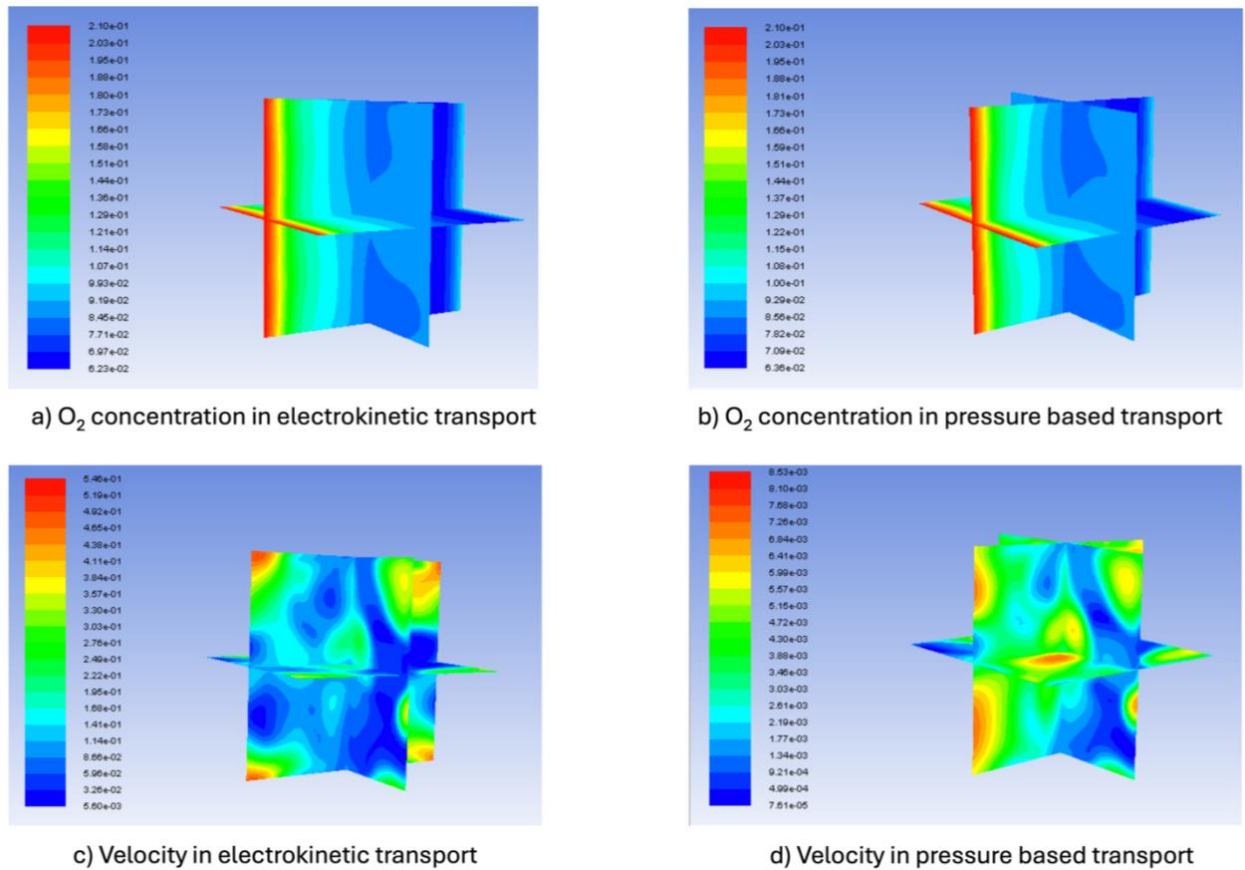

Fig 7: Comparison of Oxygen ($O_2$) concentraion and velocity profiles in electrokinetic and pressure driven flow.